\documentclass[11pt]{article}
\usepackage{amssymb}
\newcommand{\equal}{\!\!\!&=&\!\!\!}

\begin{document}
\abovedisplayshortskip 12pt
\belowdisplayshortskip 12pt
\abovedisplayskip 12pt
\belowdisplayskip 12pt
\baselineskip=17pt
\title{{\bf The Bi-Hamiltonian Structure of the Short Pulse Equation}} 
\author{J. C. Brunelli  \\
\\
Departamento de F\'\i sica, CFM\\
Universidade Federal de Santa Catarina\\
Campus Universit\'{a}rio, Trindade, C.P. 476\\
CEP 88040-900\\
Florian\'{o}polis, SC, Brazil\\
}
\date{}
\maketitle

\begin{center}
{ \bf Abstract}
\end{center}

We prove the integrability of the short pulse equation derived recently by Sch\"afer and Wayne from a hamiltonian point of view. We give its bi-hamiltonian structure and show how the recursion operator defined by the hamiltonian operators is connected with the one obtained by Sakovich and Sakovich. An alternative zero-curvature formulation is also given.
\bigskip



\newpage

The cubic nonlinear Schr\"odinger (NLS) equation is used to describe the propagation of optical pulses in nonlinear media.
Extending the work of Altermnan and Rauch \cite{alterman} the short pulse (SP) equation
\begin{equation}
u_{xt}=u+{1\over 6}\,\left( u^3\right)_{xx}\;,\label{spe}
\end{equation}
was derived by Sch\"afer and Wayne \cite{schafer} as an alternative equation to the NLS equation to describe the evolution of very short optical pulses in nonlinear media. Also, they made numeric studies that proved that as the pulse length gets short the SP equation is a much better approximation to the solution of Maxwell's equation  than the NLS equation \cite{chung}. These results support the integrability of the SP equation based on numerical techniques. In an interesting paper Sakovich and Sakovich \cite{sakovich} studied the integrability of the SP equation from a Lax or zero-curvature point of view. They  found that the  equation (\ref{spe}) possesses a
Lax pair of the Wadati–-Konno–-Ichikawa type \cite{wadati} and that through a chain of transformations this equation can be related to the sine-Gordon equation. This result sets the equation (\ref{spe}) as ultrashort pulse integrable alternative to the NLS equation. Also, using cyclic basis techniques these authors obtained the following recursion operator
\begin{equation}
{\overline R}=\partial^2{1\over{u_{xx}}}\partial {({1+u_x^2})^{1/2}}\partial^{-1}{u_{xx}\over({1+u_x^2})^{3/2}}\;.\label{rss}
\end{equation}

It is well known that the integrability of a nonlinear equation can be approached from a Lax representation. Using a spectral transform method analytic solutions can in principle be obtained. However, these calculations turn out to be very complex and an alternative approach to integrability is desirable. An algebraic approach based on a hamiltonian formulation for infinite dimensional dynamical systems is very powerful and was intensely developed over the last several years after the seminal works of Gardner \cite{gardner}, Zakharov and Faddeev \cite{zakharov} and Magri \cite{magri}. 
In this letter we will show that the SP  equation  is bi-hamiltonian. These two compatible hamiltonian operators allow us to establish the integrability of the SP equation from a hamiltonian point of view without the introduction of B\"acklund transformations or the use of Lax equations. As it is well known \cite{magri} from these structures we can construct a recursion operator that yields a hierarchy of conserved equations and charges. For the SP equation its  bi-hamiltonian structure will result in a recursion operator whose inverse is exactly (\ref{rss}). A recursion operator (bi-hamiltonian structure) can also provide insights on a Lax or zero curvature formulation for an integrable system. Therefore, we will also obtain an alternative zero-curvature formulation for the SP equation which will provide in a natural way the hierarchy of charges and equations associated with it.

The equation (\ref{spe}) studied in Ref. \cite{sakovich} is related to the original form used in \cite{schafer,chung}
by  scale transformations of the dependent and independent
variables, but in order to obtain a hamiltonian description for the SP equation let us integrate (\ref{spe}) once with respect to $x$ to obtain the nonlocal equation
\begin{equation}
u_{t}=(\partial^{-1}u)+{1\over 2}\,u^2u_x\;.\label{shortpulse}
\end{equation}
Nonlocal equations, as well hierarchies  of equations, of this sort have been studied extensively in the last years \cite{brunelli1}.

Introducing the Clebsch potential
\begin{equation}
u=\phi_x\;,\label{transf}
\end{equation}
the equation (\ref{shortpulse}) can be obtained from a variational principle, $\delta\int dtdx\,{\cal L}$, with the first order lagrangian density
\begin{equation}
{\cal L}={1\over 2}\phi_t \phi_x - {1\over24} \phi_x^4 + {1\over 2}\phi^2\;.\label{lagrangian}
\end{equation}
Using the Dirac's theory of constraints \cite{dirac} we can obtain the hamiltonian and the hamiltonian operator associated with (\ref{lagrangian})
\begin{eqnarray}
{\cal D}_1\equal\partial\;,\nonumber\\
\noalign{\vskip 5pt} 
H_2 \equal \int dx\left[{1\over24}u^4-{1\over
2}(\partial^{-1}u)^2\right]\;.\label{first}
\end{eqnarray}
It can be easily checked that $H_2$ is conserved by the SP equation. In this way the SP equation in the form (\ref{shortpulse}) can be written in hamiltonian form as
\[
u_t={\cal D}_1{\delta H_2\over\delta u}\;.
\]

As it is well known a system can be shown to be integrable if it is bi-hamiltonian \cite{magri,olver,blaszak}, i.e., the system has a hamiltonian description with two distinct hamiltonian structures that are compatible. In fact, as can be easily checked, the SP equation can also be written as
\[
u_t={\cal D}_2{\delta H_1\over\delta u}\;,
\]
with
\begin{eqnarray}
{\cal D}_2\equal\partial^{-1}+u_x\partial^{-1}u_x\;,\nonumber\\
\noalign{\vskip 5pt} 
H_1 \equal {1\over2}\int dx \,u^2\;.\label{second}
\end{eqnarray}
Again, $H_1$ is conserved by the SP equation. Also, the skew symmetry of this hamiltonian structure is manifest. The proof of Jacobi identity for this structure as well the compatibility with the hamiltonian structure in (\ref{first}) can be determined through the method of prolongation \cite{olver} in a straightforward way. Therefore, the bi-hamiltonian structures (\ref{first}) and (\ref{second}) also guarantees the integrability of the SP equation from a hamiltonian point of view and also provide us with a natural recursion operator defined by
\begin{equation}
R={\cal D}_2{\cal D}_1^{-1}\;\label{r}
\end{equation}
or
\begin{equation}
R=\left(1+u_x^2-u_x\,\partial^{-1}u_{xx}\right)\partial^{-2}\;.\label{rf}
\end{equation}
It can immediately be checked that the inverse of $R$ is exactly $\overline R$ obtained by Sakovich and Sacovich \cite{sakovich}. 

From the recursion operator (\ref{r}) a whole hierarchy of equations and conserved charges can be constructed and the details will be published elsewhere \cite{brunelli0}. However, since the system is bi-hamiltonian the SP equations and the other equations of the hierarchy will have the general form
\begin{equation}
u_t=\left({\cal D}_2 X\right)-\lambda\left({\cal D}_1 X\right)\;,\label{eqhierarchy}
\end{equation}
where $\lambda$ represents a  spectral parameter and $X$ is an arbitrary function of the dynamical variable as well as of the spectral parameter. This equation can alternatively  be obtained from the zero curvature condition (see \cite{dasroy} and references therein)
\begin{equation}
\partial_t\mathbb{A}_1-\partial_x\mathbb{A}_0-[\mathbb{A}_0,\mathbb{A}_1]=0\;.\label{zerocurvature}
\end{equation}
From the SP equation the canonical dimensions of the variables are given by $[u]=-1$, $[x]=-1$, and $[t]=1$. Also, from (\ref{eqhierarchy}) we have $[X]=-1$ and $[\lambda]=-2$. Let us choose, taking into account (\ref{zerocurvature}), the gauge fields with matrices
\begin{equation}
\mathbb{A}_{0} = \left(\begin{array}{ll}
{[\ ]=-1} & {[\ ]=1}\\
\noalign{\vskip 10pt}%
{[\ ]=-3} & {[\ ]=-1}\\
\end{array}\right)\;,\quad\mathbb{A}_{1} = \left(\begin{array}{ll}
{[\ ]=1} & {[\ ]=3}\\
\noalign{\vskip 10pt}%
{[\ ]=-1} & {[\ ]=1}\\
\end{array}\right)\label{a0a1}
\end{equation}
where $[\ ]$ represents the dimensionality of the matrix element. Choosing the gauge fields as
\begin{eqnarray}
\mathbb{A}_0\equal\left(%
\begin{array}{cc}
-\lambda^{-1/2}(\partial^{-1}u)  &  \lambda^{-1/2}\\
  \noalign{\vskip 12pt}
 \displaystyle(\partial^{-2}X)+u(\partial^{-1}u_xX)- (\partial^{-1}uu_xX)-\lambda^{-1/2}(\partial^{-1}u)^2-\lambda X \qquad & \lambda^{-1/2}(\partial^{-1}u) \\
\end{array}%
\right)\nonumber\;,\\\noalign{\vskip 15pt}
\mathbb{A}_1\equal \left(%
\begin{array}{cc}
 \lambda^{-1/2} & 0 \\
  \noalign{\vskip 12pt}
  u & \lambda^{-1/2} \\
\end{array}%
\right)\;,\label{Alocal}
\end{eqnarray}
the zero curvature condition (\ref{zerocurvature}) indeed yields the dynamical equations in (\ref{eqhierarchy})
\begin{equation}
u_t= \left(\partial^{-1} + u_{x}\partial^{-1} u_x\right)X-\lambda\,\partial X\;.\label{equation}
\end{equation}
Dynamical variables are independent of the spectral parameter, therefore, the function $X$ must depend on $\lambda\/$ for this equation to be meaningful. Making  a Taylor expansion in $\lambda$ of the form
\begin{equation}
X(\lambda,u)=\sum_{j=0}^{n}\lambda^{n-j}X_j(u)\;,\label{expansion}
\end{equation}
with 
\begin{equation}
X_0={\displaystyle u_{xx}\over{\sqrt{(1+u_x^2)^3}}}\;,\label{initial1}
\end{equation}
and substituting (\ref{expansion}) and (\ref{initial1}) into (\ref{equation}), we obtain
\begin{eqnarray}
{\cal D}_{2} X_{j}(u)\equal{\cal D}_{1} X_{j+1}(u)\;,\qquad j = 0,1,2,\dots ,n-1\;,\nonumber\\
\noalign{\vskip 10pt}%
u_{t_n}\equal {\cal D}_2X_n\;.\label{zerorecursion}
\end{eqnarray}
If we identify
\begin{equation}
X_{j} = \frac{\delta H_{j}}{\delta u}\;,\label{identification}
\end{equation}
then (\ref{zerorecursion}) gives the dynamical equations of the hierarchy (for any $n$), the two hamiltonian structures of the system as well as the recursion relations between the conserved charges of the hierarchy. For completeness, the first flows are
\begin{eqnarray}
\begin{array}{l}
\displaystyle u_{t_{0}}=u_x\;,\\
\noalign{\vspace{15pt}}
\displaystyle u_{t_{1}}=(\partial^{-1}u)+{1\over 2}\,u^2u_x\;,\\
\noalign{\vspace{15pt}}
\displaystyle u_{t_{2}}=(\partial^{-3}u)+{1\over 6}(\partial^{-1}u^3)+ u_x\left(\partial^{-1}\left(u_x\left(\partial^{-2}u\right)\right)\right)+{1\over 24}\,u^4u_x\;,\\
\quad\,\,\,\vdots\;.
\end{array}\label{nonlocalflow}
\end{eqnarray}
The $n=0$ flow is the chiral equation and the $n=1$ flow is just the SP equation (\ref{shortpulse}). The conserved charges for the hierarchy can, of course, be determined recursively from
(\ref{zerorecursion}) and give the infinite set of  conserved hamiltonians
\begin{eqnarray}
H_0\equal-\int dx\,\sqrt{1+u_x^2}\;,\nonumber\\\noalign{\vskip 7pt}
H_1\equal{1\over2}\int dx\,u^2\;,\nonumber\\\noalign{\vskip 7pt}
H_2\equal\int dx\left[{1\over24}u^4-{1\over
2}(\partial^{-1}u)^2\right]\;,\nonumber\\\noalign{\vskip 7pt}
H_3\equal\int  dx\,\left[{1\over 720}u^6+{\frac{1}{2}}(\partial^{-2}u)^2+{\frac{1}{6}}(\partial^{-2}u^3)u-
{\frac{1}{4}}(\partial^{-1}u)^2u^2\right]\;,\nonumber\\\noalign{\vskip 7pt}
&\vdots&\;.\label{nonlocal}
\end{eqnarray}

Our zero-curvature formulation (\ref{Alocal}) for $X=u$ and $\lambda=0$ is nonlocal and yields the SP equation in the form (\ref{shortpulse}). Sakovich and Sakovich have also obtained a zero-curvature for the SP equation written in the local form (\ref{spe}). Both zero-curvatures are related by some sort of gauge transformation.

In this letter we have supported the integrability of the short pulse equation  through its bi-hamiltonian structure. We related our work with the previous work of Sakovich and Sakovich \cite{sakovich} comparing their recursion operator with ours. A zero-curvature formulation providing a hierarchy of conserved charges and equations, containing the SP equation, was given. Following the results of \cite{brunelli1} this hierarchy can be extended to include other equations such as the elastic beams under tension equation \cite{ichikawa} and the details can be found in \cite{brunelli0}.

\section*{Acknowledgments}

This work was supported by CNPq (Brazil).

\end{document}